\documentclass[english]{article}
\usepackage{amsfonts,amsmath,amsthm,amscd,amssymb,latexsym}

\theoremstyle{plain}
\newtheorem{thm}{Theorem}[section]

\newtheorem{lem}[thm]{Lemma}
\newtheorem{prop}[thm]{Proposition}
\newtheorem{rem}[thm]{Remark}
\theoremstyle{definition}
\theoremstyle{remark}
\numberwithin{equation}{section}
\newcommand{\keywords}{\textbf{Key words and phrases: }\medskip}
\newcommand{\subjclass}{\textbf{Math. Subj. Clas.: }\medskip}

\begin{document}
\title{\textbf{A gauge-invariant discrete analog of \\ the Yang-Mills
equations \\ on a double complex} }
\author{\textbf{Volodymyr Sushch} \\
{ Technical University of Koszalin} \\
 { Sniadeckich 2, 75-453 Koszalin, Poland} \\
 { sushch@lew.tu.koszalin.pl} }

\date{}
\maketitle
\begin{abstract}
 An intrinsically defined gauge-invariant discrete model of the
 Yang-Mills equations on a combinatorial analog
 of $\Bbb{R}^4$ is constructed. We develop several algebraic
 structures on the matrix-valued cochains (discrete forms)
 that are analogs of objects in differential geometry. We define a combinatorial
 Hodge star operator based on the
use of a double complex construction. Difference
 self-dual and anti-self-dual  equations will be given. In the last section we
discus the question of generalizing our constructions  to the case
of a 4-dimensional combinatorial sphere.
\end{abstract}

\keywords{Yang-Mills equations, gauge invariance, difference
equations}

 \subjclass  {81T13, 39A12}

\section{Introduction}
In this paper we construct a gauge-invariant discrete model of the
Yang-Mills equations and study combinatorial analogs of some
objects in differential geometry, namely the Hodge star operator,
the self-dual and anti-self-dual equations. We define these
structures on a combinatorial analog of $\Bbb{R}^4$ based on the
use of a double complex.

Using the approach first introduced by Dezin \cite{Dezin}, in
\cite{S1, S2}  we consider gauge-invariant discrete models of the
Yang-Mills equations in $\Bbb{R}^n$ and in Min\-kowski space. Our
approach based also on some constructions from \cite{Dezin1},
where certain $2$-dimensional models connected with the Yang-Mills
equations are discussed.  In \cite{S1} the combinatorial Hodge
star operator $\ast$ is defined using both an inner product on
discrete matrix-valued forms (cochains) and Poincar$\acute{e}$
duality but the operation $(\ast)^2$ is equivalent to a shift with
corresponding sign. This is one of the main distinctive features
of the formalism \cite{Dezin} as compared to the continual case,
where the operator $(\ast)^2$ is either an involution or
antiinvolution.

In with paper we introduce a combinatorial object, namely a double
complex, in which the discrete Hodge star operator is defined in
such way that  $(\ast)^2=\pm Id$. At the same time we consider
discrete forms, the product $\cup$ on cochains (discrete analog of
the exterior product) and the coboundary operator $d^c$ (discrete
analog of the exterior differential operator) similarly as in
\cite{S1, S2}.

There is another approach presented in Dodziuk's paper \cite{D}.
In \cite{D, DP} the authors using an embedding of simplicial
cochains into differential forms (due to \cite{W}) show that a
combinatorial Laplacian on the cochains provide a good
approximation of the smooth Laplacian on a closed Riemannian
manifold.  Using the techniques \cite{D}, Wilson \cite{WS} defines
a combinatorial star operator on the simplicial cochains of a
triangulated Riemannian  manifold and proves its convergence to
the smooth star operator. Other related results on the subject can
be found in \cite{Bak, Kom, SS}.

In section 4 we construct a discrete analog of the Yang-Mills
equations on the cochains of the double complex. We try to be as
close to continual Yang-Mills theory as possible. We'll define the
discrete  Yang-Mills equations using both a geometric structure of
the object and a gauge invariance of these equations.

 A large number of paper in the physical literature have been devoted to
discretization of gauge theories (see, for example, \cite{Amb, A,
ASS, BR, Se} and the references therein). Discrete version of
Yang-Mills theories using lattices and graphs, as well as their
applications to finite dimensional versions of gauge theories,
have been studied in \cite{Nish, Shab, Frol}.  Some other
interesting results on gauge invariant lattice models of
Yang-Mills actions using the geometry of finite groups can be
found in \cite{Cast}.

It is well known that in $4$-dimensional non-abelian gauge theory
the self-dual and anti-self-dual connections are the most
important extrema of the Yang-Mills action. In section 5 we study
discrete analogs of the self-dual and anti-self-dual equations on
the double complex and show that some interesting relations
amongst the curvature form and its self-dual and anti-self-dual
parts, that hold in the continual theory, also hold in the
combinatorial case. We also describe difference self-dual
equations as a system of nonlinear matrix equations.

In section 6 we consider a combinatorial construction which is
homeomorphic to a $4$-dimensional sphere. The technique
introduced, namely the combinatorial Hodge star operator on the
cochains of the double complex,  allows us to describe a discrete
model of the Yang-Mills equations on a combinatorial
$4$-dimensional sphere. Using the same approach a discrete analog
of the Laplacian on a combinatorial $2$-dimensional sphere is
considered in \cite{S3}.  It is interesting to study problems like
this since the question concerning the global approximations
throughout the surface of a ball has not been study enough.
\section{Continual setting}
In this section we briefly recall some well known definitions of
smooth Yang-Mills theory (see, for example, \cite{NS}). Let $M$ be
a smooth oriented Riemannian manifold. Consider the trivial bundle
${P=M\times SU(2)}$. Let $T^\ast P$ be the cotangent bundle of $P$
and let $(x,g)$, $x\in M$, $g\in SU(2)$, be local coordinates of
the bundle $P$. It is known \cite{NS} that a connection can be
shown to arise from a certain 1-form $\omega\in T^\ast P$, where
$\omega$ is required to have values in the Lie algebra $su(2)$ of
the Lie group $SU(2)$ . This form is given by
\begin{equation} \label{2.1}
\omega=g^{-1}dg+g^{-1}Ag.
\end{equation}
The connection 1-form $A$ is defined by
\begin{equation} \label{2.2}
A=\sum_{\alpha,\mu}A_{\mu}^{\alpha}(x)\lambda_{\alpha} dx^\mu,
\end{equation}
where $\lambda_\alpha$ is a basis of $su(2)$ and
$A_{\mu}^{\alpha}(x)$ is a smooth function for any $\mu, \alpha$.
 Here we take $\lambda_\alpha=\frac{\sigma_\alpha}{2i}$,  where
$\sigma_\alpha, \ \alpha=1,2,3$,  are the standard Pauli matrices
and  $i$ is the imaginary unit.

Let the coordinates of $P$ change (locally) from $(x,g)$ to
$(x^\prime, g^\prime)$. Let us only make a change of fibre
coordinates, i.e. $x^\prime=x$ and $g^\prime$ is given by
\begin{equation} \label{2.3}
g^\prime=hg, \qquad h\in SU(2).
\end{equation}
 Under the change of coordinates (\ref{2.3}) the 1-form
$\omega$ induces a certain transformation law for the connection
form $A$. Suppose that the form (\ref{2.1}) is invariant under
transformations (\ref{2.3}), i.e.
 \begin{equation*} \label{2.4}
(g^\prime)^{-1}dg^\prime+(g^\prime)^{-1}A^\prime g^\prime=
g^{-1}dg+g^{-1}Ag.
 \end{equation*}
From with  we obtain
\begin{equation} \label{2.5}
A^\prime=hdh^{-1}+hAh^{-1}.
\end{equation}
 This is the transformation law of the connection form $A$. It is what is called in  Yang-Mills theories
 the gauge transformation law.

We define the curvature 2-form $F$ in the following way
\begin{equation} \label{2.6}
 F=dA+A\wedge A.
 \end{equation}
  Under the gauge transformation  (\ref{2.3}) the curvature $F$ changes as
  follows
\begin{equation} \label{2.7}
 F^\prime=hFh^{-1}.
 \end{equation}

Define the covariant exterior differential operator $d_A$ by
\begin{equation} \label{2.8}
d_A\Omega=d\Omega+A\wedge \Omega+(-1)^{r+1}\Omega\wedge A,
\end{equation} where $\Omega$ is a $su(2)$-valued $r$-form.

Then the Yang-Mills equations can be written as
\begin{equation} \label{2.9} d_A F=0,
\end{equation}
\begin{equation} \label{2.10}  d_A\ast F=0,
\end{equation}
 where $\ast$ is the Hodge star operator. Equation (\ref{2.9}) is known as the Bianchi identity.
 Let $\Phi, \Psi$ be $su(2)$-valued  $r$-forms on $M$. The inner product is defined by
\begin{equation} \label{2.11}
 (\Phi, \Psi)=-tr\int_{M}\Phi\wedge\ast\Psi,
 \end{equation}
  where $tr$ is the trace operator.

Let $M$ be  $4$-dimensional. The Yang-Mills action S can be
expressed in terms of the 2-forms $F$ and $\ast F$ as
\begin{equation} \label{2.12}
 S=-tr\int_{M}F\wedge\ast F.
 \end{equation}
 Equations (\ref{2.9}), (\ref{2.10}) are the Euler-Lagrange
 equations for the extrema of $S$. In $4$-dimensional Yang-Mills
 theories the following equations
 \begin{equation} \label{2.13}
 F=\ast F, \qquad F=-\ast F
 \end{equation}
 are called self-dual and anti-self-dual respectively.
 Solutions of (\ref{2.13}) -- the self-dual and anti-self-dual
 connections -- are called also instantons and antiinstantons \cite{FU}.
 It is known that the self-dual and anti-self-dual connections are the
 most important minima of the action $S$.
 \section {Double complex}
 Let the tensor product  $C(4)=C\otimes C\otimes C\otimes C$
of an  1-dimensional complex $C$ be a combinatorial model of
Euclidean space $\Bbb{R}^4$ (see for details \cite{Dezin, S2}).
The 1-dimensional complex $C$ is defined in the following way. Let
$C^0$ denotes the real linear space of 0-dimensional chains
generated by basis elements $x_\kappa$ (points), $\kappa\in
\Bbb{Z}$. It is convenient to introduce the shift operators
$\tau,\sigma$ in the set of indices by $$ \tau\kappa=\kappa+1,
\qquad \sigma\kappa=\kappa-1. $$ We denote the open interval
$(x_\kappa, \ x_{\tau\kappa)}$ by $e_\kappa$. We'll regards the
set $\{e_\kappa\}$ as a set of basis elements of the real linear
space $C^1$  of 1-dimensional chains. Then the 1-dimensional
complex (combinatorial real line) is the direct sum of the
introduced spaces $C=C^0\oplus C^1$. The boundary operator
$\partial$ on the basis elements of $C$ is given by
 \begin{equation} \label{3.1}
 \partial x_\kappa=0, \qquad  \partial
e_\kappa=x_{\tau\kappa}-x_\kappa.
 \end{equation}
 The definition is extended to arbitrary chains by linearity.

Multiplying the basis elements $x_\kappa, e_\kappa$ in various way
we obtain basis elements of $C(4)$. Let $s_k^{(p)}$, where
$k=(k_1,k_2,k_3,k_4), \ k_i\in\Bbb Z,$ be an arbitrary basis
element of $C(4)$. We suppose that the superscript $(p)$ contains
the whole requisite information about the number and places of the
1-dimensional "components" $e_\kappa$  in $s_k^{(p)}$.
 For example, 1-dimensional basis elements
of $C(4)$ can be written as
\begin{align}\label{3.2}\notag
 e_k^1&= e_{k_1}\otimes
x_{k_2}\otimes x_{k_3}\otimes x_{k_4}, \qquad e_k^2=
x_{k_1}\otimes e_{k_2}\otimes x_{k_3}\otimes x_{k_4}, \\ e_k^3&=
x_{k_1}\otimes x_{k_2}\otimes e_{k_3}\otimes x_{k_4}, \qquad
e_k^4=x_{k_1}\otimes x_{k_2}\otimes x_{k_3}\otimes e_{k_4}
\end{align}
 and for the 2-dimensional basis elements
$\varepsilon_k^{ij}$ we have
\begin{align}\label{3.2a}\notag
\varepsilon_k^{12}&=e_{k_1}\otimes e_{k_2}\otimes x_{k_3}\otimes
e_{k_4}, \qquad \varepsilon_k^{23}=x_{k_1}\otimes e_{k_2}\otimes
e_{k_3}\otimes x_{k_4}, \\ \notag
\varepsilon_k^{13}&=e_{k_1}\otimes x_{k_2}\otimes e_{k_3}\otimes
x_{k_4}, \qquad \varepsilon_k^{24}=x_{k_1}\otimes e_{k_2}\otimes
x_{k_3}\otimes e_{k_4},
 \\ \varepsilon_k^{14}&=e_{k_1}\otimes x_{k_2}\otimes
x_{k_3}\otimes e_{k_4}, \qquad \varepsilon_k^{34}=x_{k_1}\otimes
x_{k_2}\otimes e_{k_3}\otimes e_{k_4}.
\end{align}
Using (\ref{3.1}), we define the boundary operator
$\partial$ on chains of $C(4)$ in the following way: if $c_p, \
c_q$ are chains of the indicated dimension, belonging to the
complexes being multiplied, then
\begin{equation}\label{3.3}
\partial(c_p\otimes c_q)=\partial c_p\otimes c_q+(-1)^pc_p\otimes\partial c_q.
\end{equation}
For example, for the basis element $\varepsilon_k^{24}$ we have
\begin{align*}
\partial\varepsilon_k^{24}&=\partial(x_{k_1}\otimes e_{k_2})\otimes
x_{k_3}\otimes e_{k_4}-x_{k_1}\otimes e_{k_2}\otimes
\partial(x_{k_3}\otimes e_{k_4}) \\ &=\partial x_{k_1}\otimes
e_{k_2}\otimes x_{k_3}\otimes e_{k_4}+x_{k_1}\otimes \partial
e_{k_2}\otimes x_{k_3}\otimes e_{k_4}\\ &- x_{k_1}\otimes
e_{k_2}\otimes \partial x_{k_3}\otimes e_{k_4}-x_{k_1}\otimes
e_{k_2}\otimes x_{k_3}\otimes \partial e_{k_4} \\ &=
x_{k_1}\otimes  x_{\tau k_2}\otimes x_{k_3}\otimes
e_{k_4}-x_{k_1}\otimes  x_{k_2}\otimes x_{k_3}\otimes e_{k_4}\\
&-x_{k_1}\otimes x_{k_2}\otimes x_{k_3}\otimes x_{\tau
k_4}+x_{k_1}\otimes x_{k_2}\otimes x_{k_3}\otimes x_{k_4}.
\end{align*}

We now describe the construction of a double complex. Together
with the complex $C(4)$ we consider its "double", namely the
complex $\tilde{C}(4)$ of exactly the same structure. Define the
one-to-one correspondence
\begin{equation}\label{3.4}
\ast : C(4)\rightarrow\tilde{C}(4), \qquad \ast : \tilde
C(4)\rightarrow C(4)
\end{equation}
in the following way. Let  $s_k^{(p)}$ be an arbitrary
$p$-dimensional basis element of $C(4)$, i.e.  the product
$s_k^{(p)}=s_{k_1}\otimes s_{k_2}\otimes s_{k_3}\otimes s_{k_4}$
contains exactly $p$ $1$-dimensional elements $e_{k_i}$ and $4-p$
\ $0$-dimensional elements  $x_{k_i}$, $p=0,1,2,3,4$, \
$k_i\in\Bbb Z.$ Then
\begin{equation}\label{3.5}
\ast : s_k^{(p)}\rightarrow\pm\tilde s_k^{(4-p)}, \qquad \ast :
\tilde s_k^{(4-p)}\rightarrow \pm s_k^{(p)},
\end{equation}
where
\begin{equation*}
 \tilde s_k^{(4-p)}=*s_{k_1}\otimes *s_{k_2}\otimes
*s_{k_3}\otimes *s_{k_4}
\end{equation*}
and $*s_{k_i}=\tilde e_{k_i}$ if $s_{k_i}=x_{k_i}$ and
$*s_{k_i}=\tilde x_{k_i}$ if $s_{k_i}=e_{k_i}.$ In the first of
mapping  (\ref{3.5}) we take "$+$" if the permutation $((p), \
(4-p))$ of $(1,2,3,4)$ is even and "$-$" if the permutation $((p),
\ (4-p))$ is odd. Recall that in symbol $(p)$ the number of
components is contained. For example, for the 2-dimensional basis
element $\varepsilon_k^{13}=e_{k_1}\otimes x_{k_2}\otimes
e_{k_3}\otimes x_{k_4}$ we have
 $\ast\varepsilon_k^{13}=-\tilde\varepsilon_k^{24}$ since the
 permutation $(1,3,2,4)$ is odd. The mapping $\ast :
\tilde s_k^{(4-p)}\rightarrow \pm s_k^{(p)}$ is defined by
analogy.
\begin{prop}Let $c_r\in C(4)$ be an $r$-dimensional chain.
Then we have
\begin{equation}\label{3.6}
\ast\ast c_r=(-1)^{r(4-r)}c_r.
\end{equation}
\end{prop}
\begin{proof} For $r=0, 4$ it is obviously. Let $r=1$, then for the 1-dimensional basis
elements (\ref{3.2}) we have
\begin{align*}
\ast e_k^1= \tilde x_{k_1}\otimes\tilde e_{k_2}\otimes \tilde
e_{k_3}\otimes \tilde e_{k_4}=\tilde e_k^{234}, \qquad \ast e_k^2=
-\tilde e_{k_1}\otimes\tilde x_{k_2}\otimes \tilde e_{k_3}\otimes
\tilde e_{k_4}=-\tilde e_k^{134}, \\ \ast e_k^3= \tilde
e_{k_1}\otimes\tilde e_{k_2}\otimes \tilde x_{k_3}\otimes \tilde
e_{k_4}=\tilde e_k^{124}, \qquad \ast e_k^4= -\tilde
e_{k_1}\otimes\tilde e_{k_2}\otimes \tilde e_{k_3}\otimes \tilde
x_{k_4}=-\tilde e_k^{123}
\end{align*}
and
\begin{align*}
\ast \tilde e_k^{123}=e_k^4, \quad \ast \tilde e_k^{124}=-e_k^3,
\quad \ast \tilde e_k^{134}=e_k^2, \quad \ast \tilde
e_k^{234}=-e_k^1.
\end{align*}
Hence $\ast\ast e_k^i=-e_k^i$ for any $i=1,2,3,4$. The case $r=3$
is similar.

Let now $\varepsilon_k^{ij}\in C(4)$ be a  2-dimensional basis
element (\ref{3.2a}) . Then
\begin{align*}
\ast\ast\varepsilon_k^{12}=\ast\tilde\varepsilon_k^{34}=\varepsilon_k^{12},
\quad
\ast\ast\varepsilon_k^{13}=-\ast\tilde\varepsilon_k^{24}=\varepsilon_k^{13},
\quad
\ast\ast\varepsilon_k^{14}=\ast\tilde\varepsilon_k^{23}=\varepsilon_k^{14},\\
\ast\ast\varepsilon_k^{23}=\ast\tilde\varepsilon_k^{14}=\varepsilon_k^{23},
\quad
\ast\ast\varepsilon_k^{24}=-\ast\tilde\varepsilon_k^{13}=\varepsilon_k^{24},
\quad
\ast\ast\varepsilon_k^{34}=\ast\tilde\varepsilon_k^{12}=\varepsilon_k^{34}.
\end{align*}
 To an arbitrary
chain $c_r$ the operation $\ast$ extends by linearity.
\end{proof}

Now we  consider a dual object of the  complex  $C(4)$. Let $K(4)$
be a cochain complex with  $gl(2,\Bbb{C})$-valued coefficients,
where  $gl(2,\Bbb{C})$ is the Lie algebra of all complex
$2\times2$ matrices. We suppose that the complex $K(4)$, which is
a conjugate of $C(4)$, has a similar structure: ${K(4)=K\otimes
K\otimes K\otimes K}$, where $K$ is a dual of the 1-dimensional
complex $C$. Basis elements of $K$ can be written as
$\{x^\kappa\}, \{e^\kappa\}$. Then an arbitrary basis element of
$K(4)$ is given by ${s^k= s^{k_1}\otimes s^{k_2}\otimes
s^{k_3}\otimes s^{k_4}}$, where $s^{k_j}$ is either $x^{k_j}$ or
$e^{k_j}$. For example, we denote the 1-, 2-dimensional basis
elements of $K(4)$ by $e_i^k$, \ $\varepsilon^k_{ij}$
respectively, cf. (\ref{3.2}), (\ref{3.2a}).

 We define the pairing operation for arbitrary
basis elements \ $\varepsilon_k\in C(4)$, \ $s^k\in K(4)$ by the
rule
\begin{equation}\label{3.7} <\varepsilon_k, as^k>=\left\{\begin{array}{l}0,\
\varepsilon_k\ne s_k\\
                            a,\ \varepsilon_k=s_k,\ a\in gl(2,\Bbb{C}).
                            \end{array}\right.
\end{equation}
 The operation (\ref{3.7}) is linearly extended to cochains. We will
call cochains forms, emphasizing their relationship with the
corresponding continual objects, differential forms.

The operation $\partial$ (\ref{3.3}) induces the dual operation
$d^c$ on $K(4)$ in the following way:
\begin{equation}\label{3.8}
<\partial\varepsilon_k, as^k>=<\varepsilon_k, ad^cs^k>.
\end{equation}
 The coboundary operator $d^c$ is an analog of the exterior
differentiation operator.

 Now we describe a cochain product on the forms of $K(4)$. See \cite{Dezin, S1, S2}
 for details. We denote this product by
 $\cup$. In terms of the homology theory this is the so-called Whitney product.
  First we introduce the $\cup$-product on the chains of the 1-dimensional complex K.
  For the basis elements of $K$ the
$\cup$-product is defined as follows $$ x^\kappa\cup
x^\kappa=x^\kappa, \quad e^\kappa\cup x^{\tau\kappa}=e^\kappa,
\quad x^\kappa\cup e^\kappa=e^\kappa, \quad \kappa\in\Bbb{Z}, $$
supposing the product to be zero in all other case. To arbitrary
forms the $\cup$-product be extended linearly. Let us introduce an
$r$-dimensional complex $K(r)$, ${r=1,2,3}$,\ in an obvious
notation. Let $s_{(p)}^k$ be an arbitrary $p$-dimensional basis
element of $K(r)$. It is convenient to write the basis  element of
$K(r+1)$ in the form   $s_{(p)}^k\otimes s^\kappa$, where
$s_{(p)}^k$ is a basis element of $K(r)$ and $s^\kappa$ is either
$e^\kappa$  or $x^\kappa$, \ $\kappa\in\Bbb{Z}$. Then, supposing
that the $\cup$-product in $K(r)$ has been defined, we introduce
it for basis elements of $K(r+1)$ by the rule
\begin{equation}\label{3.9} (s^k_{(p)}\otimes
s^\kappa)\cup(s^k_{(q)}\otimes s^\mu)= Q(\kappa,q)(s^k_{(p)}\cup
s^k_{(q)})\otimes(s^\kappa\cup s^\mu),
\end{equation}
 where  the signum
function $Q(\kappa, q)$ is equal to $-1$ if the dimension of both
elements $s^\kappa$, $s_{(q)}^k$ is odd and to $+1$ otherwise (see
\cite{Dezin}). The extension of the $\cup$-product to arbitrary
forms of $K(r+1)$ is linear.  Note that the coefficients of forms
multiply as matrices.

\begin{prop}{Let $\varphi$ and $\psi$ be arbitrary forms of $K(4)$.
Then
\begin{equation}\label{3.10}  d^c(\varphi\cup\psi)=d^c\varphi\cup\psi+(-1)^p\varphi\cup
d^c\psi,
\end{equation} where  $p$ \ is the dimension of a form
$\varphi$.}
\end{prop}

The proof of Proposition 3.2 is totally analogous to one in
\cite[p.~147]{Dezin} for the case of discrete forms with real
coefficients.

The complex of the cochains $\tilde K(4)$ over the double complex
$\tilde C(4)$, with the operator $d^c$ defined in it by
(\ref{3.8}), has the same structure as  $K(4)$. The operation
(\ref{3.8}) induces the respective mapping
\begin{equation*}
\ast : K(4)\rightarrow\tilde{K}(4), \qquad \ast : \tilde
K(4)\rightarrow K(4)
\end{equation*}
by the rule:
\begin{equation}\label{3.11} <\tilde
c, \ *\varphi>=<*\tilde c, \ \varphi>, \qquad <c, \ *\tilde
\psi>=<*c, \ \tilde\psi>,
\end{equation}
where $c\in C(4), \ \tilde c\in\tilde C(4), \ \varphi\in K(4), \
\tilde\psi\in \tilde K(4)$. It is obviously that Proposition~3.1
is true for any $r$-dimensional cochain $c^r\in K(4)$. So we have
$$\ast\ast\varphi=(-1)^{r(4-r)}\varphi$$ for any discrete $r$-form
$\varphi$ on $K(4)$ and note that the same relation holds in the
continual case.

Let $V\subset C(4)$ be a "domain" of the complex $C(4)$. We define
its as follows
\begin{equation}\label{3.12}
 V=\sum_kV_k, \qquad k=(k_1,k_2,k_3,k_4),\quad k_i=1,2, ...,N_i,
\end{equation}
where $V_k= e_{k_1}\otimes e_{k_2}\otimes e_{k_3}\otimes e_{k_4}$
is a 4-dimensional basis element of $C(4)$. Let $s_k^{(p)}$ be a
$p$-dimensional basis element of  $C(4)$. We set
\begin{equation}\label{3.13}
 V_p=\sum_k\sum_{(p)}s_k^{(p)}\otimes\ast s_k^{(p)},
\end{equation}
where the subscripts $k_i,\ i=1,2,3,4$,  run the set of values
indicated in (\ref{3.12}). For example,
\begin{equation*}
 V_1=\sum_k\sum_{i=1}^4 e_k^i\otimes\ast e_k^i=\sum_k(e_k^1\otimes\tilde e_k^{234}-
 e_k^2\otimes\tilde e_k^{134}+e_k^3\otimes\tilde e_k^{124}-e_k^4\otimes\tilde
 e_k^{123}).
 \end{equation*}
 Let $K(V)$ denotes  $K(4)$ restricted to $V$ and let
 \begin{equation*}
 \mathbb{V}=\sum_{p=0}^4 V_p.
 \end{equation*}
Consider the following discrete $p$-forms
\begin{equation*}
\varphi= \sum_k \sum_{(p)}\varphi_k^{(p)}s_{(p)}^k,  \qquad
\varphi^*= \sum_k \sum_{(p)}\big(\varphi_k^{(p)}\big)^*s_{(p)}^k,
 \end{equation*}
where $\varphi_k^{(p)}\in gl(2,\Bbb{C})$ and
$\big(\varphi_k^{(p)}\big)^*$ denotes the conjugate transpose of
the matrix $\varphi_k^{(p)}$, i. e.
$\big(\varphi_k^{(p)}\big)^*=\big(\overline{\varphi}_k^{(p)}\big)^T$.

   For any $p$-forms $\varphi, \psi\in K(V)$ we define the inner
 product  $(\ , \ )_V$ by
 \begin{align}\label{3.14}\notag
 (\varphi ,\ \psi)_V&=tr<\mathbb{V}, \ \varphi\otimes\ast\psi^* >=tr<V_p, \ \varphi\otimes\ast\psi^* >\\
 &=tr\sum_k\sum_{(p)}
 <s_k^{(p)},\
\varphi ><\ast s_k^{(p)}, \ \ast\psi^* >.
\end{align}
Using (\ref{3.5}), (\ref{3.7}), it is easy to check that
\begin{equation}\label{3.15}
(\varphi ,\ \psi)_V=tr\sum_k
\sum_{(p)}\varphi_k^{(p)}\big(\psi_k^{(p)}\big)^*,
\end{equation}
where $\varphi_k^{(p)}, \big(\psi_k^{(p)}\big)^*\in gl(2,\Bbb{C})$
are components of $\varphi,\psi^*\in K(V)$.

Note that for $su(2)$-valued $p$-forms on $V$ (cf. (\ref{2.11}))
Relation~(\ref{3.15}) can be rewritten as follows
\begin{equation*}
 (\varphi ,\ \psi)_V=-tr<\mathbb{V}, \ \varphi\otimes\ast\psi >=-tr\sum_k\sum_{(p)}
 \varphi_k^{(p)}\psi_k^{(p)}.
\end{equation*}

The inner product makes it possible to define the adjoint of
$d^c$, denoted $\delta^c$.
\begin{prop} For any $(p-1)$-form $\varphi$ and $p$-form $\omega$ we have
\begin{equation}\label{3.16}
(d^c\varphi, \ \omega)_V=tr<\partial \mathbb{V}, \
\varphi\otimes\ast\omega^*>+( \varphi, \ \delta^c\omega)_V,
\end{equation} where
 \begin{equation}\label{3.17}
 \delta^c=(-1)^{p}\ast^{-1}d^c\ast
\end{equation}
and $\ast\ast^{-1}=Id$.
\end{prop}
\begin{proof} The proof is a computation. From the definition
(\ref{3.8}) it follows that  (\ref{3.3})  induces the similar
relation for the coboundary operator $d^c$ on forms:
\begin{equation*}
d^c(\varphi\otimes\ast\omega)=d^c\varphi\otimes\ast\omega+(-1)^{p-1}\varphi\otimes
d^c(\ast\omega).
\end{equation*}
Using this, we compute
\begin{align*}
(d^c\varphi, \ \omega)_V&=tr<\mathbb{V}, \
d^c\varphi\otimes\ast\omega^*>=tr<V_p, \
d^c\varphi\otimes\ast\omega^*>\\&= tr<\mathbb{V}, \
d^c(\varphi\otimes\ast\omega^*)>-(-1)^{p-1}tr<V_{p-1}, \
\varphi\otimes d^c(\ast\omega^*)>\\&=tr <\partial \mathbb{V}, \
\varphi\otimes\ast\omega^*>+(-1)^{p}tr<\mathbb{V}, \
\varphi\otimes \ast(\ast^{-1}d^c\ast\omega)^*>.
 \end{align*}
 It immediately follows (\ref{3.16}).
\end{proof}
Relation (\ref{3.16}) is a discrete analog of the Green formula.
It should be noted that from (\ref{3.6}) we have:
$$\ast^{-1}=(-1)^{p(4-p)}\ast.$$

\section{Discrete Yang-Mills equations}
In this section we'll construct a discrete model of the Yang-Mills
equations (\ref{2.9}), (\ref{2.10}) using the double complex
introduced above. Let $A\in K(4)$ be a discrete 1-form. We can
write $A$ as
\begin{equation}\label{4.1}
A=\sum_k\sum_{i=1}^4A_k^ie_i^k,
\end{equation}
where \ $A_k^i\in su(2)$ \ and \ $e_i^k$ \ is an 1-dimensional
basis element of \ $K(4)$, \ $k=(k_1,k_2,k_3,k_4), \ k_i\in\Bbb
Z.$ Suppose that the $su(2)$-valued 1-form (\ref{4.1}) is a
discrete analog of the connection form (\ref{2.2}).

Let us introduce some discrete 0-dimensional form with
coefficients belonging to the Lie group $SU(2)$. We put
\begin{equation}\label{4.2}
h=\sum_kh_kx^k,
\end{equation}
where $h_k\in SU(2)$ and $x^k=x^{k_1}\otimes x^{k_2}\otimes
x^{k_3}\otimes x^{k_4}$ is a 0-dimensional basis element of
$K(4)$.

The discrete analog of the  transformations (\ref{2.3}),
(\ref{2.5}) are defined to be
\begin{equation}\label{4.3}
 g^\prime=h\cup g, \qquad A^\prime=h\cup
d^ch^{-1}+h\cup A\cup h^{-1},
\end{equation}
where $h, h^{-1}, g$ are 0-forms of the type (\ref{4.2}) and
$h^{-1}$ denotes the form with inverse coefficients (inverse
matrices). We'll call this transformation a gauge transformation
for the discrete model.
\begin{rem} The set of the  0-forms (\ref{4.2}) is a group with respect to the
$\cup$-product.
\end{rem}
It is obviously, since by definition of the $\cup$-product for the
0-forms $h, \ g$ we have
\begin{equation*}
h\cup
g=\Big(\sum_kh_kx^k\Big)\cup\Big(\sum_kg_kx^k\Big)=\sum_kh_kg_kx^k,
\end{equation*}
where $h_k, \ g_k$ are multiplied as matrices.

 The discrete curvature form $F$ is defined by
\begin{equation}\label{4.4}
F=d^cA+A\cup A.
\end{equation}
 The 2-form $F\in K(4)$ we can write also as follows
\begin{equation}\label{4.5}
F=\sum_k\sum_{i<j} F_k^{ij}\varepsilon_{ij}^k,
 \end{equation}
 where \ $ F_k^{ij}\in gl(2,\Bbb{C})$, \  $ \varepsilon_{ij}^k$ \ is a
  2-dimensional  basis elements of \ $K(4)$ \ and \ $1\leq i,j\leq4$, \ $k=(k_1,k_2,k_3,k_4)$, \ $k_i\in\Bbb Z$.

   Let us introduce for convenient  the shifts
operator $\tau_i$ and $\sigma_i$ as $$\tau_ik=(k_1,...\tau
 k_i,...k_4), \qquad
 \sigma_ik=(k_1,...\sigma k_i,...k_4).$$
  Similarly, we denote by
 $\tau_{ij}$ ($\sigma_{ij}$) the operator shifting to the right (to the left) two differ components of $k=(k_1,k_2,k_3,k_4)$.
 For example, $$\tau_{12}k=(\tau k_1,\tau k_2,k_3,k_4), \qquad \sigma_{14}k=(\sigma k_1,k_2,k_3,\sigma k_4).$$
Combining (\ref{4.4}) and (\ref{4.5}) and using (\ref{3.7}) --
(\ref{3.9}), we obtain
 \begin{equation}\label{4.6}
 F_k^{ij}=\Delta_{k_i}A_k^j-\Delta_{k_j}A_k^i+A_k^iA_{\tau_ik}^j-
 A_k^jA_{\tau_jk}^i,
 \end{equation}
 where $\Delta_{k_i}A_k^j=A_{\tau_ik}^j-A_k^j$.
 \begin{rem}In the continual case the curvature form $F$
 (\ref{2.6}) takes values in the algebra $su(2)$ for any $su(2)$-valued connection form $A$.
 Unfortunately, it is not true in the discrete case because, generally speaking, the components
 $A_k^iA_{\tau_ik}^j- A_k^jA_{\tau_jk}^i$ of the form $A\cup A$ (see (\ref{4.6})) do not belong to $su(2)$.
 \end{rem}

 It is easy to check that the combinatorial Bianchi identity:
 \begin{equation}\label{4.7}
 d^cF=A\cup F-F\cup A
\end{equation}
holds for the discrete curvature form (\ref{4.4}) (cf.
(\ref{2.9})).
 The discrete analog of the exterior covariant
differentiation operator (\ref{2.8}) is defined by
\begin{equation}\label{4.8}
d_A^c\Omega=d^c\Omega+A\cup\Omega+(-1)^{r+1}\Omega\cup A,
\end{equation}
 where $\Omega$ is an arbitrary $r$-form of $K(4)$.
\begin{thm} Under the gauge transformation (\ref{4.3}) the curvature form
(\ref{4.4}) changes as
\begin{equation}\label{4.9}
F^\prime=h\cup F\cup h^{-1}.
 \end{equation}
\end{thm}
\begin{proof}
The proof is similar to that of Proposition 2, \cite{S2}.
\end{proof}

Let us introduce the following operation  $$\tilde\iota: K(4)
\rightarrow \tilde K(4), \qquad \tilde\iota: \tilde K(4)
\rightarrow K(4)$$ by setting
\begin{equation}\label{4.10}
 \tilde\iota s_{(p)}^k= \tilde s_{(p)}^k, \qquad \tilde\iota\tilde s_{(p)}^k=  s_{(p)}^k,
\end{equation}
where $s_{(p)}^k$ and $\tilde s_{(p)}^k$ are  basis elements of
$K(4)$ and $\tilde K(4) $.  So, for a $p$-form  $\varphi\in K(4)$
we have \ $\tilde\iota\varphi=\tilde\varphi$. \ Recall that the
coefficients of $\tilde\varphi\in \tilde K(4)$ and  $\varphi\in
K(4)$ are the same.
\begin{prop} The following hold
\begin{align}\label{4.11}
\tilde\iota^2=Id, \quad \tilde\iota\ast&=\ast\tilde\iota, \quad
\tilde\iota d^c=d^c\tilde\iota,\\ \notag
\tilde\iota(\varphi\cup\psi)&=\tilde\iota\varphi\cup\tilde\iota\psi,
\end{align}
where $\varphi, \ \psi\in K(4)$.
\end{prop}
\begin{proof}
The proof immediately follows from definitions of the
corresponding operations.
\end{proof}

The discrete analog of Equation (\ref{2.10}) can be written as
\begin{equation}\label{4.13}
  d_A^c\ast\tilde\iota F=0.
\end{equation}
Using (\ref{4.8}), we have
\begin{equation}\label{4.14}
d_A^c\ast\tilde\iota F= d^c\ast\tilde\iota F+A\cup\ast\tilde\iota
F-\ast\tilde\iota F\cup A.
\end{equation}
\begin{lem} Let $h$ be a discrete 0-form. Then for an arbitrary $p$-form $f\in
K(4)$ we have
\begin{equation}\label{4.15}
 \tilde\iota\ast(h\cup f)=h\cup\tilde\iota\ast f.
 \end{equation}
\end{lem}

\begin{proof} Applying (\ref{4.10}), the proof is analogous to the proof of Lemma~1 in
\cite{S2}.
\end{proof}

\begin{lem} Let $f\in K(4)$ be a  2-form. We have
\begin{equation}\label{4.16}\tilde\iota\ast(f\cup h)=\tilde\iota\ast f\cup h
 \end{equation}
   if and only if the coefficients of a 0-form $h$ satisfy the
following conditions
\begin{equation}\label{4.17}
h_{\tau_{12}k}=h_{\tau_{34}k}, \qquad
h_{\tau_{13}k}=h_{\tau_{24}k}, \qquad
h_{\tau_{14}k}=h_{\tau_{23}k}
\end{equation}
 for all $k=(k_1,k_2,k_3,k_4)$, $k_i\in\Bbb Z$.
\end{lem}
\begin{proof} The proof is computational. See also the proof of
Lemma~2 in \cite{S2}. Using (\ref{3.9}) and (\ref{3.5}), we
compute

\begin{equation*}
  f\cup
h=\sum_k\sum_{i<j}f_k^{ij}h_{\tau_{ij}k}\varepsilon_{ij}^k,
\end{equation*}
and
\begin{align*}
  \ast f=\sum_k(f_k^{12}\tilde\varepsilon_{34}^k-f_k^{13}\tilde\varepsilon_{24}^k+
f_k^{14}\tilde\varepsilon_{23}^k+f_k^{23}\tilde\varepsilon_{14}^k-
f_k^{24}\tilde\varepsilon_{13}^k+f_k^{34}\tilde\varepsilon_{12}^k),
\end{align*}
 where $\varepsilon_{ij}^k$ is a 2-dimensional basis element
of $K(4)$.
 Then, by the definition of $\tilde\iota$, we obtain
 \begin{align*}
 \tilde\iota\ast(f\cup h)=
\sum_k(f_k^{12}h_{\tau_{12}k}\varepsilon_{34}^k-f_k^{13}h_{\tau_{13}k}\varepsilon_{24}^k&+
f_k^{14}h_{\tau_{14}k}\varepsilon_{23}^k\\+f_k^{23}h_{\tau_{23}k}\varepsilon_{14}^k-
f_k^{24}h_{\tau_{24}k}\varepsilon_{13}^k&+f_k^{34}h_{\tau_{34}k}\varepsilon_{12}^k).
\end{align*}
On the other hand, we have
\begin{align*}
 \tilde\iota\ast f\cup h=
\sum_k(f_k^{12}h_{\tau_{34}k}\varepsilon_{34}^k-f_k^{13}h_{\tau_{24}k}\varepsilon_{24}^k&+
f_k^{14}h_{\tau_{23}k}\varepsilon_{23}^k\\+f_k^{23}h_{\tau_{14}k}\varepsilon_{14}^k-
f_k^{24}h_{\tau_{13}k}\varepsilon_{13}^k&+f_k^{34}h_{\tau_{12}k}\varepsilon_{12}^k).
\end{align*}
Combining the last two expressions with one another, we conclude
that (\ref{4.16}) implies (\ref{4.17}) and vice versa.
\end{proof}

It should be noted that  the set of 0-forms (\ref{4.2}), which
satisfy Conditions (\ref{4.17}), is a group under $\cup$-product
(see Remark~4.1).
\begin{thm} Under Conditions (\ref{4.17}) the discrete Yang-Mills
equation (\ref{4.13}) is gauge invariant.
\end{thm}
\begin{proof}
The proof is analogous to the proof of Theorem~1 in \cite{S2}. By
Theorem~4.2 and Lemmas~4.4, 4.5, we have
\begin{equation*}
\tilde\iota\ast F^{\prime}=\tilde\iota\ast(h\cup F\cup
h^{-1})=h\cup\tilde\iota\ast F\cup h^{-1}.
\end{equation*}
Note that $h^{-1}$ also satisfies Conditions (\ref{4.17}). Using
(\ref{3.17}) we compute
\begin{equation*}
d^c\tilde\iota\ast F^{\prime}=d^ch\cup\tilde\iota\ast F\cup
h^{-1}+h\cup d^c\tilde\iota\ast F\cup h^{-1}+h\cup \tilde\iota\ast
F\cup d^ch^{-1}.
\end{equation*}
Since $d^ch\cup h^{-1}=-d^ch\cup h^{-1}$ and taking into account
(\ref{4.3}) and (\ref{4.9}), we get
\begin{equation*}
 A^\prime\cup\tilde\iota\ast
F^\prime=-d^ch\cup\tilde\iota\ast F\cup h^{-1}+ h\cup
A\cup\tilde\iota\ast F\cup h^{-1}
\end{equation*} and
\begin{equation*} \tilde\iota\ast F^\prime\cup A^\prime=h\cup\tilde\iota\ast F\cup d^ch^{-1}+
h\cup\tilde\iota\ast F\cup A\cup h^{-1}.
\end{equation*}
Putting the last  three expressions in  (\ref{4.14}), one obtains:
\begin{align*}
  d^c_{A^\prime}\tilde\iota\ast F^\prime & =h\cup d^c\tilde\iota\ast F\cup h^{-1}+h\cup A\cup\tilde\iota\ast
F\cup h^{-1} \\
 & -h\cup\tilde\iota\ast F\cup A\cup h^{-1} =h\cup d_A^c\tilde\iota\ast F\cup
h^{-1}.
\end{align*}
Thus, if $d^c_{A}\tilde\iota\ast F=0$, then
$d^c_{A^\prime}\tilde\iota\ast F^\prime=0$.

\end{proof}
\section{Difference self-dual and anti-self-dual equations}
The discrete analog of Equations (\ref{2.13}) is defined by
\begin{equation}\label{5.1}
 F=\tilde\iota\ast F, \qquad F=-\tilde\iota\ast F,
\end{equation}
where $F$ is the discrete curvature form (\ref{4.4}). Using
(\ref{4.5}), by the definitions of $\tilde\iota$ and $\ast$, the
first equation (self-dual) of (\ref{5.1}) can be rewritten as
follows
\begin{equation}\label{5.2}
 F_k^{12}=F_k^{34}, \qquad F_k^{13}=-F_k^{24}, \qquad
 F_k^{14}=F_k^{23}.
\end{equation}
We call these equations difference self-dual equations. Using
(\ref{4.6}), we obtain
\begin{equation*}
 \Delta_{k_1}A_k^2-\Delta_{k_2}A_k^1+A_k^1A_{\tau_1k}^2-
 A_k^2A_{\tau_2k}^1=\Delta_{k_3}A_k^4-\Delta_{k_4}A_k^3+A_k^3A_{\tau_3k}^4-
 A_k^4A_{\tau_4k}^3,
 \end{equation*}
 \begin{equation*}
 \Delta_{k_1}A_k^3-\Delta_{k_3}A_k^1+A_k^1A_{\tau_1k}^3-
 A_k^3A_{\tau_3k}^1=\Delta_{k_4}A_k^2-\Delta_{k_2}A_k^4-A_k^2A_{\tau_2k}^4+
 A_k^4A_{\tau_4k}^2,
 \end{equation*}
 \begin{equation*}
 \Delta_{k_1}A_k^4-\Delta_{k_4}A_k^1+A_k^1A_{\tau_1k}^4-
 A_k^4A_{\tau_4k}^1=\Delta_{k_2}A_k^3-\Delta_{k_3}A_k^2+A_k^2A_{\tau_2k}^3-
 A_k^3A_{\tau_3k}^2.
 \end{equation*}
 Recall that $A_k^i\in su(2)$ is a component of the discrete connection
 1-form (\ref{4.1}).

 Obviously, changing the sign on the right hand side of Equations (\ref{5.2}),
 we obtain the difference anti-self-dual equations.

 As in the continual case (see, for example, \cite{NS}),
  we can decompose our arbitrary discrete 2-form $F$ into its
  self-dual and anti-self-dual parts as follows
\begin{equation*}
  F=F^++F^-,
\end{equation*}
where $F^+=\frac{1}{2}(F+\tilde\iota\ast F)$  and
$F^-=\frac{1}{2}(F-\tilde\iota\ast F)$. The form $F^+$ is
self-dual, i.e. $F^+=\tilde\iota\ast F^+$. Indeed, using
Proposition~3.1 and (\ref{4.11}), we compute
\begin{equation*}
  \tilde\iota\ast F^+=\tilde\iota\ast \frac{1}{2}(F+\tilde\iota\ast
  F)=\frac{1}{2}(\tilde\iota\ast F+\tilde\iota^2\ast^2 F)=\frac{1}{2}(\tilde\iota\ast F+
  F)=F^+.
\end{equation*}
Similarly,
\begin{equation*}
  \tilde\iota\ast F^-=\tilde\iota\ast \frac{1}{2}(F-\tilde\iota\ast
  F)=\frac{1}{2}(\tilde\iota\ast F-\tilde\iota^2\ast^2 F)=\frac{1}{2}(\tilde\iota\ast F-
  F)=-F^-.
\end{equation*}
So, $F^-$ is anti-self-dual.

Let $\| \ \|_V$ denote the norm on $K(V)$  generated by the inner
 product (\ref{3.14}). Then a discrete analog of the Yang-Mills
 action (\ref{2.12}) can be written  as
\begin{equation*}
  S=\| F \|^2_V=(F, \ F)_V=tr<V_2, \ F\otimes\ast F^*>.
\end{equation*}
See also (\ref{3.12})--(\ref{3.15}).
\begin{thm}
For the discrete curvature form (\ref{4.4}) we have
\begin{equation}\label{5.3}
 \| F \|^2_V=\| F^+ \|^2_V+\| F^- \|^2_V.
\end{equation}
\end{thm}
\begin{proof} By definition (\ref{3.14}) we have
\begin{align*}
  \| F \|^2_V &=\| F^++F^- \|^2_V  \\
   &=tr<V_2, \ F^+\otimes\ast (F^+)^*>+tr<V_2, \ F^-\otimes\ast (F^-)^*>\\
   &+tr<V_2, \ F^+\otimes\ast (F^-)^*>+tr<V_2, \ F^-\otimes\ast (F^+)^*>\\
   &=\| F^+ \|^2_V+\| F^- \|^2_V+(F^+, \ F^-)_V+(F^-, \ F^+)_V.
\end{align*}
Denote the components of $F^+$, \ $F^-$ by $(F_k^{ij})^+$, \
$(F_k^{ij})^-$ respectively. For $(F_k^{ij})^+$ we have
(\ref{5.2}) and for $(F_k^{ij})^-$ we can write the following
relations
\begin{equation*}
 (F_k^{12})^-=-(F_k^{34})^-, \qquad (F_k^{13})^-=(F_k^{24})^-, \qquad
 (F_k^{14})^-=-(F_k^{23})^-.
\end{equation*}
Then, using (\ref{3.15}), we obtain
\begin{align*}
 (F^+, \ F^-)_V&=tr\sum_k\sum_{i<j}(F_k^{ij})^+\big[(F_k^{ij})^-\big]^*=-tr\sum_k\sum_{i<j}(F_k^{ij})^+\big[(F_k^{ij})^-\big]^* \\
   &=-(F^+, \ F^-)_V.
\end{align*}
 Thus,
 $(F^+, \ F^-)_V=0$.  Similarly, we have $(F^-, \ F^+)_V=0$.
 \end{proof}
It should be noted that in the continual case Relation (\ref{5.3})
implies that the self-dual and anti-self-dual connections
(solutions of (\ref{2.13})) are always absolute minima of the
action $S$ (see \cite{NS}).

\section{Combinatorial model of the 4-sphere}
In this section we discus the question of generalizing our
constructions introduced above to the case of a 4-dimensional
complex which is the boundary of a 5-dimensional domain. Note that
constructions used in \cite{S1,S2}, namely the operation $\ast$,
are inappropriate to this case. It is convenient to employ here a
construction based on the use of the double complex. We will use a
standard technique ("gluing of the double") that turns a manifold
with boundary into a manifold without boundary.

Let $V\in C(4)$ be a "domain" in the form (\ref{3.12}). Together
with $V\in C(4)$ we introduce its counterpart $\hat V\in C(4)$.
Considering now $V, \ \hat V$ to be two distinct domains and
identifying the respective elements of the boundary, we obtain the
4-dimensional combinatorial manifold $M=V\bigcup\hat V$ which is
homeomorphic to the 4-dimensional sphere. Let $s_k^{(p)}$ be a
basis element of $C(V)$. Denote by $\hat s_k^{(p)}$ the
corresponding basis element of $\hat C(V)$. The "gluing"
conditions of $V$ and $\hat V$ are defined by
\begin{align}\label{6.1}\notag
 s_{k_1...0...k_4}^{(p)}=\hat s_{k_1...N_i...k_4}^{(p)},\qquad s_{k_1...\tau N_i...k_4}^{(p)} &= \hat
s_{k_1...1...k_4}^{(p)},\\ \hat s_{k_1...0...k_4}^{(p)}=
s_{k_1...N_i...k_4}^{(p)},\qquad \hat s_{k_1...\tau
N_i...k_4}^{(p)} &= s_{k_1...1...k_4}^{(p)},
\end{align}
where $0\leq k_i\leq N_i$, \ see (\ref{3.12}). On the other hand,
a new combinatorial object, namely the complex $C(M)$, is defined
by Conditions (6.1). The boundary operator $\partial$ on $C(M)$ is
given by (\ref{3.3}). We call the complex $C(M)$ a combinatorial
4-dimensional sphere. As in section 3, we introduce the dual
complex $K(M)$. No essential modifications are needed to carry out
constructions, considered in  $K(4)$, in the complex $K(M)$. An
arbitrary $p$-form $\varphi\in K(M)$ can be written as
\begin{equation*}
  \varphi=\sum_k\sum_{(p)}(\varphi_k^{(p)}s_{(p)}^k+\hat\varphi_k^{(p)}\hat
  s_{(p)}^k),
\end{equation*}
where $\varphi_k^{(p)}, \ \hat\varphi_k^{(p)}\in gl(2,\Bbb{C})$ \
and $ s_{(p)}^k\in K(V), \ \hat s_{(p)}^k\in K(\hat V)$ are
corresponding basis elements, $k=(k_1, k_2, k_3, k_4)$,
 $k_i=1,2, ...,N_i$. Due to the definition (\ref{3.12}),
 Conditions (6.1) imply the following  conditions for the form
 $\varphi$:
\begin{align}\label{6.2}\notag
 \varphi_{k_1...0...k_4}^{(p)}=\hat \varphi_{k_1...N_i...k_4}^{(p)},\qquad \varphi_{k_1...\tau N_i...k_4}^{(p)} &= \hat
\varphi_{k_1...1...k_4}^{(p)},\\
\hat\varphi_{k_1...0...k_4}^{(p)}=
\varphi_{k_1...N_i...k_4}^{(p)},\qquad \hat\varphi_{k_1...\tau
N_i...k_4}^{(p)} &= \varphi_{k_1...1...k_4}^{(p)}.
\end{align}
Recall that the components  $\varphi_{k_1...0...k_4}^{(p)}$, \
$\varphi_{k_1...\tau N_i...k_4}^{(p)}$ appear when we consider the
coboundary operator $d^c$ and its applications.

It is obvious that the double complex construction extends to the
complex $C(M)$ (or $K(M)$). The star operation $\ast$ on $K(M)$ is
also defined by (\ref{3.12}). So, for any $p$-forms $\varphi,
\psi\in K(M)$ the inner product can be written as
\begin{equation*}
 (\varphi ,\ \psi)_M=tr<V_p, \ \varphi\otimes\ast\psi^* >+tr<\hat V_p, \ \hat\varphi\otimes\ast\hat\psi^* >
\end{equation*}
or
\begin{equation*}
(\varphi ,\ \psi)_M=tr\sum_k
\sum_{(p)}\big(\varphi_k^{(p)}\big(\psi_k^{(p)}\big)^*+\hat\varphi_k^{(p)}\big(\hat\psi_k^{(p)}\big)^*\big).
\end{equation*}

The discrete connection 1-form over $C(M)$ is defined by
\begin{equation*}
 A=\sum_k\sum_{i=1}^4 (A_k^ie_i^k+\hat A_k^i\hat
e_i^k),
\end{equation*}
where $A_k^i, \hat A_k^i\in su(2)$, \ $e_i^k, \hat e_i^k$  are
1-dimensional basis elements of $K(M)$,  $k=(k_i, k_2, k_3, k_4),
\  k_i=1,2, ..., N_i$. Similarly, the discrete curvature 2-form
(\ref{4.4}) can be written as
\begin{equation*}
 F=\sum_k\sum_{i<j}
(F_k^{ij}\varepsilon_{ij}^k+\hat F_k^{ij}\hat\varepsilon_{ij}^k),
\end{equation*}
where $F_k^{ij}, \hat F_k^{ij}\in gl(2,\Bbb{C})$, \
$\varepsilon_{ij}^k, \hat\varepsilon_{ij}^k$ are  2-dimensional
basis elements of $K(M)$. The components $A_k^i, \hat A_k^i,
F_k^{ij}, \hat F_k^{ij}$ satisfy Conditions (6.2).

It is easy to check that all constructions from Sections 4, 5
carry out in the complex $K(M)$. Thus, we can write the discrete
Yang-Mills equations in the form (\ref{4.7}), (\ref{4.13}) and
Theorem~4.7  holds on the combinatorial sphere $C(M)$. In $K(M)$
the difference self-dual and anti-self-dual equations (\ref{5.2})
are completed by the same equations for the components $\hat
F_k^{ij}$. So, under Conditions~6.2 we obtain the
finite-dimensional system of matrices equations.

\end{document}